\documentclass[prx, 10pt, twocolumn, floatfix, superscriptaddress, aps, longbibliography, nofootinbib]{revtex4-2}

\usepackage{times, mathrsfs, amsmath, amsfonts, graphics, graphicx, cancel, color, amsthm, bbm, mathtools, amssymb, physics, tikz, quantikz}

\usepackage{amsmath,amsthm,amsfonts,graphicx,xcolor,times,xfrac,booktabs, mathtools,enumitem,xr,subcaption,amssymb,bbm,verbatim,appendix,placeins, physics}
\usepackage[unicode=true,bookmarks=true,bookmarksnumbered=false,bookmarksopen=false,breaklinks=false,pdfborder={0 0 1}, backref=false,colorlinks=true,citecolor=red]{hyperref}
\usepackage{bm}
\usepackage{dcolumn,algorithm,algpseudocode}

\makeatletter
\theoremstyle{plain}
\newtheorem{thm}{\protect\theoremname}
\theoremstyle{plain}

\theoremstyle{plain}

\theoremstyle{remark}
\newtheorem*{rem*}{\protect\remarkname}
\theoremstyle{plain}

\theoremstyle{plain}

\theoremstyle{definition}

\theoremstyle{plain}
\newtheorem*{thm*}{\protect\theoremname}
\theoremstyle{plain}
\newtheorem*{lem*}{\protect\lemmaname}

\providecommand{\propositionname}{Proposition}
\providecommand{\theoremname}{Theorem}
\providecommand{\lemmaname}{Lemma}
\providecommand{\remarkname}{Remark}
\providecommand{\conjecturename}{Conjecture}
\providecommand{\definitionname}{Definition}
\providecommand{\corollaryname}{Corollary}
\allowdisplaybreaks

\def\bra#1{\langle{#1}\vert}
\def\ket#1{\vert{#1}\rangle}
\def\braket#1{\langle{#1}\rangle}

\def\BraVert{e.g.,roup\,\mid\,\bgroup}

\def\ketbra#1#2{\vert{#1}\rangle\!\langle{#2}\vert}

\begin{document}
\title{Optical Fourier Architecture for Universal Nonlinear Functions}

\author{Martin F. X. Mauser}
\email{martin.mauser@univie.ac.at}
\affiliation{University of Vienna, Faculty of Physics, Vienna Center for Quantum
Science and Technology (VCQ), Boltzmanngasse 5, Vienna 1090, Austria}
\affiliation{University of Vienna, Vienna Doctoral School in Physics,  Boltzmanngasse 5, Vienna 1090, Austria}
\author{Joshua Morris}
\email{joshua.morris@univie.ac.at}
\affiliation{University of Vienna, Faculty of Physics, Vienna Center for Quantum
Science and Technology (VCQ), Boltzmanngasse 5, Vienna 1090, Austria}
\affiliation{University of Vienna, Vienna Doctoral School in Physics,  Boltzmanngasse 5, Vienna 1090, Austria}
\author{Sara Galatro}
\affiliation{University of Vienna, Faculty of Physics, Vienna Center for Quantum
Science and Technology (VCQ), Boltzmanngasse 5, Vienna 1090, Austria}
\affiliation{University of Vienna, Vienna Doctoral School in Physics,  Boltzmanngasse 5, Vienna 1090, Austria}
\author{Philip Walther}
\affiliation{University of Vienna, Faculty of Physics, Vienna Center for Quantum
Science and Technology (VCQ), Boltzmanngasse 5, Vienna 1090, Austria}
\affiliation{Institute for Quantum Optics and Quantum Information Sciences (IQOQI), Austrian Academy of Sciences, Boltzmanngasse 3, Vienna 1090, Austria}
\affiliation{QUBO Technology GmbH, Vienna 1090, Austria}
\author{Borivoje Daki\'{c}}
\affiliation{University of Vienna, Faculty of Physics, Vienna Center for Quantum
Science and Technology (VCQ), Boltzmanngasse 5, Vienna 1090, Austria}
\affiliation{Institute for Quantum Optics and Quantum Information Sciences (IQOQI), Austrian Academy of Sciences, Boltzmanngasse 3, Vienna 1090, Austria}
\affiliation{QUBO Technology GmbH, Vienna 1090, Austria}
\date{\today}

\begin{abstract}
We introduce an exact algebraic architecture that evaluates an arbitrary finite Fourier series using a two-mode ($2 \times 2$) linear optical circuit, with the only tunable components being single-mode phase shifters encoding the function argument. We prove that such a circuit must exist for every Fourier series and derive an analytical method for its construction based on spectral factorisation. The resulting optical system exhibits an $\mathcal{O}(N)$ depth for an $N$-harmonic expansion, executing function evaluations in the passive optical time-of-flight. Finally, we validate our claims numerically, demonstrating that even for sequences with thousands of Fourier terms, our proposed circuit construction correctly synthesises continuous and discontinuous nonlinear functions. Our architecture thus provides a universal, deterministic foundation for single-variable nonlinear optical computing on integrated photonic platforms.
\end{abstract}

\maketitle

\section{Introduction}

The exceptional growth of modern computing workloads, driven predominantly by deep neural network inference, the training of large language models, and high-bandwidth information processing, is rapidly colliding with a fundamental physical and economic barrier: the escalating energy, cooling, and supporting infrastructure costs of conventional digital electronics \cite{luccioniPower2023,strubellEnergy2020}. 

Standard information processing architectures using digital logic face severe limitations regarding clock frequencies, information routing, and thermodynamic efficiency \cite{markovLimitsFundamentalLimits2014,kimAddressingInterconnectChallenges2024, nahmiasPhotonicMultiplyAccumulateOperations2020} in ways that are deeply tied to their mode of operation and cannot easily be separated from. As our energy infrastructure struggles to keep pace with the power demands of current and future artificial intelligence workloads, optical computing has re-emerged as a compelling alternative \cite{mcmahonPhysicsOpticalComputing2023}. Coherent optical processors are already being introduced as high-throughput accelerators for linear matrix-vector operations, offering massive parallelism, sub-nanosecond propagation latencies, and superior energy efficiency \cite{mcmahonPhysicsOpticalComputing2023, nahmiasPhotonicMultiplyAccumulateOperations2020,shenDeepLearningCoherent2017,nechipurenkoPhotonicAcceleratorsAI2026}.

A central challenge in optical computing remains: how to embed nonlinear transformations into a dynamical process that is strictly linear without relying on resource-intensive, latency-heavy conversions between optical and electronic domains \cite{mcmahonPhysicsOpticalComputing2023,nahmiasPhotonicMultiplyAccumulateOperations2020,chenDeepLearningCoherent2023}. We demonstrate how arbitrary nonlinear functions may be evaluated using a linear optical system via their representation as a truncated Fourier series. Our architecture is ideally suited for Photonic Integrated Circuits (PICs) \cite{wangIntegratedPhotonicQuantum2020,barzaghiLowloss24modeLaserwritten2025}, executing function evaluations in the passive time-of-flight required for light to traverse the circuit once the input argument is encoded onto physical phase shifters. 

Previously, compiling parameterised optical circuits for function synthesis has conventionally relied on nonlinear numerical optimisation or heuristic search routines that suffer from local minima and convergence bottlenecks \cite{Mauser:2025cho, karoobyProgrammableNonlinearFunction2026}. In contrast, we provide an exact algebraic construction that proves both the universal existence of the requisite optical circuit and the method for its construction. By solving the underlying spectral factorisation problem exactly, our protocol gives an analytical solution to the global nonlinear program, which enables a near-trivial computation of a sequence of physical beam-splitter and phase-shifter settings in a single pass with a computational complexity that scales as $\mathcal{O}(N)$ for an $N$-harmonic Fourier series approximation to the target nonlinear function. 

Fundamentally, this method is inspired by the operational structure of data re-uploading \cite{perez-salinasDataReuploadingUniversal2020}, a framework developed for quantum machine learning and experimentally demonstrated on PICs \cite{Mauser:2025cho}, while the algebraic structure echoes the techniques used in Quantum Signal Processing (QSP) \cite{lowMethodologyResonantEquiangular2016,lowOptimalHamiltonianSimulation2017}. We emphasise that although our algorithm draws direct inspiration from quantum information science, the resulting computation operates entirely on classical coherent optical amplitudes. This eliminates the need for challenging nonlinear optical materials, single-photon sources, or measurement-induced nonlinearities, opening an accessible, deterministic route to programmable nonlinear operations on near-term integrated photonics. 

In what follows, we will establish the exact correspondence between a finite Fourier series and two-mode linear optical interferometers. By phrasing the solution in terms of the spectral factorisation of a  nonlinear program, our results provide a complete foundation for embedding highly nontrivial functions into optical systems. Beyond all-optical neural network activations and signal processing, this work bridges universal function approximation and photonic computing hardware.

\section{Optical Data Encoding}
Consider an arbitrary one-dimensional target function $f(x)$ that is defined over the compact interval $x \in [a,b]$. In general, the Fourier series of any square-integrable function $f \in L^2 ([a, b])$ converges pointwise almost everywhere \cite{carlesonConvergenceGrowthPartial1966}. Broadly speaking, Fourier representations establish universal approximation capabilities, where the truncated Fourier series and derived trigonometric networks are capable of learning any continuous mapping to an arbitrary precision $\epsilon > 0$ given sufficiently many terms \cite{gallantThereExistsNeural1988,hornikMultilayerFeedforwardNetworks1989,Mauser:2025cho}. This universality makes Fourier series an attractive and mathematically rigorous framework for a broad class of nonlinear operations in modern computation, with concrete examples ranging from activation functions required for artificial intelligence \cite{ngom2021} to Fourier-based polynomial arithmetic in post-quantum cryptography \cite{zeng2024}.

Specifically, over the domain $x \in [a,b]$, any continuos target function can be approximated within bounded pointwise error (and any $f \in L^2$ pointwise almost everywhere) $|f(x) - \hat{f}(x)| \leq \epsilon$, by a truncated Fourier series supporting harmonics up to cutoff order $N$:
\begin{equation}\label{eq:fourier_series}
    \hat{f}(x) = \sum_{n=-N}^N a_n e^{i n \pi x / \tau},
\end{equation}
for a chosen half-period $\tau$. Mapping the continuous input variable $x$ onto the complex unit circle via
\begin{equation}\label{eq:z_variable}
    z := e^{i \pi x / \tau}, \qquad |z| = 1,
\end{equation}
the expansion in Eq.~\eqref{eq:fourier_series} is recasted as a Laurent polynomial $\hat{f}(z) = \sum_{n=-N}^N a_n z^n$ evaluated on the unit circle.

\begin{figure}[h]
    \centering
    \includegraphics[width=0.49\textwidth]{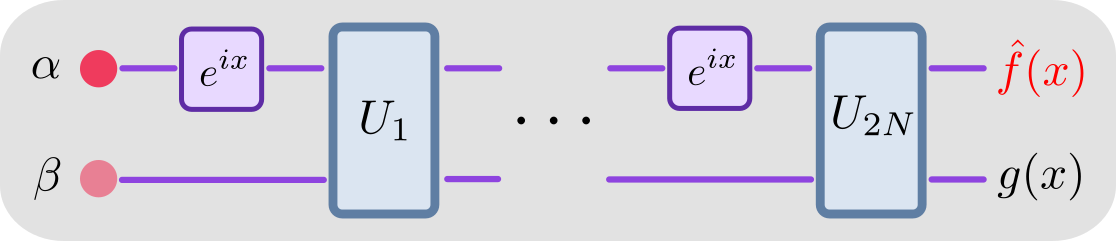}
    \caption{\textbf{Optical circuit ansatz.} A $2\times 2$ linear optical circuit acts on a fixed input state and evaluates the $N$-harmonic Fourier series $\hat{f}(x)$ for input $x$, encoded by $2N$ repeating tunable single-mode phase shifters that specify $x$ and fixed MZIs. }
    \label{fig:figure_1}
\end{figure}

The connection to linear optics in this form is particularly natural as illustrated in Figure \ref{fig:figure_1}. Rather than encoding the data parameter $x$ into the input optical state, it is instead inserted directly into the phase of a tunable single-mode phase shifter:
\begin{equation}\label{eq:phase_gate}
    T(z)=z\ketbra{0}{0} + \ketbra{1}{1} = \begin{pmatrix}
        z & 0 \\ 0 & 1
    \end{pmatrix}.
\end{equation}
This represents an instance of higher-order optical computation, where inputs are encoded as gates rather than states. Each phase shifter implements multiplication by a factor $z=e^{i\pi x/\tau}$ and hence repeated appearances of the same element with intervening static unitary operations generate higher integer powers of $z$ and consequently the relevant Fourier harmonics present in $\hat{f}(x)$. The optical circuit itself remains strictly linear with respect to the input amplitudes, but an \textit{effective} nonlinearity arises in terms of the circuit transformation and the encoded parameter $x$. The circuit construction may thus be viewed as the embedding of a nonlinear computation within a linear process acting on spatial optical modes.

In this setting, we aim to evaluate a Fourier series such that the optical amplitude in a fixed output mode remains proportional to $\hat{f}(z)$ when driven by a fixed input state. More precisely, for the target $\hat{f}(z)$, we seek the implementation  of the form
\begin{equation}
\begin{pmatrix}
    \hat{f}(z)\\ g(z)
\end{pmatrix}=z^{-N} U_{2N} T(z) U_{2N-1} T(z) \cdots U_1 T(z)\ket{\phi}
\end{equation}
via $2N$ unitaries $U_k$ acting on a fixed input $\ket{\phi}=(\alpha, \beta)^T$, where the global phase $z^{-N}$ is physically irrelevant. Because linear optical networks are strictly unitary, the target $\hat{f}(z)$ must be accompanied by an auxiliary polynomial $g(z)=\sum_kb_kz^k$, with coefficients $b_k$ to be determined, such that the total norm is conserved for all $x$. This gives a strict invariance constraint on the output state of the two-mode circuit $F(z)=(\hat{f}(z), g(z))^T$:
\begin{equation}
    \label{eq:UnitCircleIdentity}
    |\hat{f}(z)|^2 + |g(z)|^2 = C, \qquad \forall |z|=1,
\end{equation}
where  $C=|\alpha|^2+|\beta|^2$ defines the total input optical power injected into the two modes (see Figure \ref{fig:figure_1}), which can always be matched via global normalisation of the target function.
Since $|g(z)|^2 > 0$, physical realisability requires the total input power to strictly bound the target output intensity:
\begin{equation}\label{eq:norm_bound}
    C > \max_{|z| = 1} |\hat{f}(z)|^2.
\end{equation}

\vfill
\vfill

To visualise the architecture, consider the minimal series $ \hat{f}(z) = a_0 + a_1 z$. The single parameter phase shifter $T(z)$ acting on a fixed input state $\ket{\phi} = (\alpha, \beta)^T$, followed by the static unitary $U_1=(u_{ij})$, produces the upper mode amplitude $\bra{0}U_1T(z)\ket{\phi} = u_{00} \alpha z + u_{01}\beta $. Choosing $u_{00}\alpha = a_1$ and $u_{01}\beta=a_0$ yields exactly $\hat{f}(z)$. Further, for $U_1$ to be unitary, the auxiliary output $g(z) = b_0 + b_1 z$ must satisfy Eq.\eqref{eq:UnitCircleIdentity}. Expanding $|\hat{f}(z)|^2 + |g(z)|^2 = C$ on the unit circle yields two balance conditions:
\begin{equation}
\label{eq:ConditionsExample2}
\vert b_0\vert^2 + \vert b_1\vert^2 = C - \vert a_0\vert^2 - \vert a_1\vert^2 \quad \text{and} \quad b_0^* b_1  = - a_0^* a_1.
\end{equation}

The second condition is equivalent to demanding orthogonality between the harmonic coefficient vectors: $\braket{\phi_0 | \phi_1} = a_0^* a_1 + b_0^* b_1 =0$, where $\ket{\phi_n} := (a_n, b_n)^T$. The pairwise cancellation of cross-terms between $\hat{f}$ and $g$ is precisely what forces orthogonality between the coefficient vectors carrying different harmonics. Normalising $\ket{\phi_0}$ and $\ket{\phi_1}$ yields an orthonormal basis of $\mathbb{C}^2$, directly defining the static unitary $U_1$ that generates the desired amplitudes without heuristic parameter searches.  

This two-term case suggests a broader, more general pattern: for an arbitrary $N$-harmonic series, finding a complementary polynomial $g(z)$ that ensures Eq.~\eqref{eq:UnitCircleIdentity} holds for all inputs requires satisfying a coupled system of $\mathcal{O}(N)$ conditions on the auxiliary coefficients $\{b_n\}$, making it far from obvious whether an exact solution exists - let alone one that can be found without resorting to heuristic optimisation.

Our primary theoretical result shows that an exact, lossless optical circuit evaluating $\hat{f}(z)$ always exists, and further, can be compiled deterministically without optimisation heuristics.  As proven in Appendix~\ref{sec:existence} (Theorem~1) the Fej\'er-Riez theorem \cite{haagerup1992} guarantees the existence of a complementary auxiliary polynomial $g(z) = \sum_{n=-N}^N b_n z^n$ satisfying Eq.~\eqref{eq:UnitCircleIdentity} for any scale constant $C$ satisfying Eq.~\eqref{eq:norm_bound}. The auxiliary polynomial coefficients $b_n$ are computed deterministically via spectral factorisation, to be more precise via Cholesky factorisation of a Toeplitz matrix (Appendix~\ref{sec:root-finding}).

Once $g(z)$ is obtained, the requirement that the total intensity in Eq.~\eqref{eq:UnitCircleIdentity} remains constant forces the coefficient of the highest power of $z$ to vanish. Because only the extreme endpoints $p$ and $q$ contribute to this outermost harmonic, their joint coefficient vectors are analytically guaranteed to be orthogonal at every layer (Appendix~\ref{sec:constrcution}), i.e. $\braket{\phi_p | \phi_q} = 0$.
This endpoint orthogonality enables an exact layer-wise reduction, recursively removing one Fourier harmonic per stage while preserving norm conservation, and consequently reconstructing unknown unitaries $U_k$. The synthesis thereby systematically decomposes the target operation into a cascade of $2N$ static unitary transformations and phase shifters.

For a target finite Fourier series $\hat{f}(z)$, the synthesis proceeds as follows:
\begin{enumerate}
    \item{Choose any constant $C> \max_{|z|=1} |\hat{f}(z)|^2$, fixing the total input power.}
    \item{Form the strictly positive polynomial $S(z) = C - |\hat{f}(z)|^2$ using the known $a_n$ coefficients and compute its spectral factorisation to determine the auxiliary polynomial series $g(z)$ (Appendices~\ref{sec:existence} and~\ref{sec:root-finding}).}
    \item{Collect the now-determined coefficient vectors $\ket{\phi_n} = (a_n,b_n)^T$ that permit an optical unitary circuit.}
    \item{Using the now guaranteed endpoint orthogonality, determine the static unitary $U_k \in SU(2)$ that removes the outermost harmonic, and update the remaining coefficient vectors (Appendix~\ref{sec:constrcution}).}
    \item{Continue until only a single constant vector remains, $(\alpha, \beta)$. The inverse of the accumulated sequence, driven by this vector as input, produces exactly $(\hat{f}(z), g(z))^T$.}
\end{enumerate}

\section{Numerical Simulations}

\begin{figure*}[t]
    \centering
    \includegraphics[width=\textwidth]{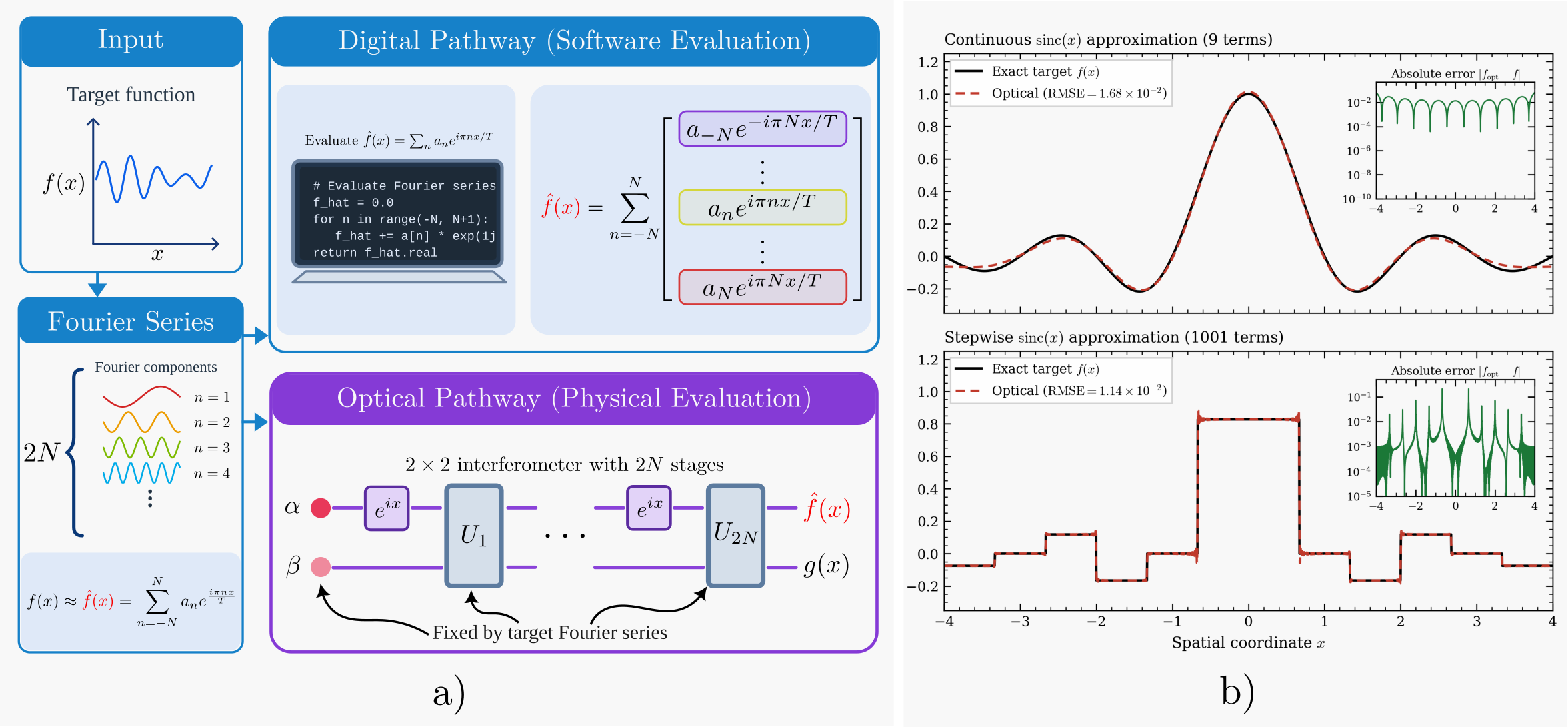}
    \caption{\label{fig:main_wide}\textbf{Optical encoding of a function}. (a) \textit{Conceptual compilation pipeline:} an arbitrary target function is approximated by a truncated Fourier series, with the quality of the approximation depending on the number of terms and the target function. Given any such series, we generate by algebraic spectral factorisation a sequence of MZIs and single-mode phase shifters that, when acting on a fixed optical input, produce the desired approximating function output $\hat{f}(x) $. (b) \textit{Simulated function approximation:} Up: the smooth target $\text{sinc} (x)$ function is faithfully reproduced by the coherent optical system using a modest circuit depth. Down: approximation of a stepwise quantized $\text{sinc} (x)$ function exhibiting discrete jump discontinuities, demonstrating numerical stability of the method as harmonic cutoff order increases.}
\end{figure*}

To demonstrate the deterministic circuit compiler and assess its numerical stability, we evaluate our method across representative nonlinear target functions. As a primary benchmark, we consider the smooth sinc function, $\text{sinc}(x) := \frac{\sin{x}}{x}$, evaluated over a compact domain (Fig.~\ref{fig:main_wide}b). 

As shown in Fig.~\ref{fig:main_wide}b (top panel), the smooth nonlinear target can be implemented using a modest circuit depth (number of MZIs and phase shifters) with a bounded error. Crucially, we emphasise that the circuit decomposition described in this paper introduces no additional algorithmic compilation error beyond the chosen Fourier truncation order. The circuit reproduces the truncated series $\hat{f}(x)$ exactly. Any residual deviations from $f(x)$ arise solely from the finite Fourier truncation and can be systematically suppressed by incorporating higher harmonics.

To rigorously probe the numerical stability, expressivity, and computational efficiency of the compiler, we next evaluate a piecewise stepped $\text{sinc} (x)$ function, obtained by rounding the continuous operation to a fixed decimal precision (Fig.~\ref{fig:main_wide}b - bottom panel).
Fixed precision rounding transforms the smooth curve into a staircase-like mapping with multiple discrete jump discontinuities. Such discontinuities are notoriously challenging for Fourier representations and are completely at odds with more traditional optical transformations. These sharp jumps induce Gibbs oscillations and slow harmonics, requiring a Fourier expansion with high cutoff order for an adequate approximation and consequently introduces strong demands on the auxiliary polynomial solver, both in terms of robustness and computational efficiency. As is demonstrated, however, the former of these is comfortably met, with the entire program run time (including setup, polynomial factorisation, circuit synthesis and plotting) remaining below five seconds on a standard Laptop CPU for a Fourier series and auxiliary polynomial and optical circuit consisting of over a thousand terms. 

A key practical advantage of our framework is the non-iterative spectral factorisation. While our numerical implementation employs the Cholesky decomposition of a banded Toeplitz matrix, alternative formulations of the Fej\'er-Riesz theorem offer distinct trade-offs depending on the harmonic cutoff $N$ and half band-width $K$:
\begin{itemize}
    \item \textit{Algebraic Root-Pairing:} For small mode cutoffs ($K \lesssim 20$), $g(z)$ can be computed in closed form by extracting the $2K$ roots of $z^K S(z)$ via companion-matrix eigenvalues \cite{golubMatrixComputations2013}. Because $S(z) > 0$, the roots pair across the unit circle as $(z_k, 1/z_k^*)$, and selecting the $K$ roots strictly inside the unit disk directly yields the canonical minimum-phase factor \cite{oppenheimDiscreteTimeSignalProcessing}. However, for higher-degree series, root extraction becomes ill-conditioned due to numerical clustering near $|z|=1$ and resulting numerical instabilities \cite{golubMatrixComputations2013}.
    \item  \textit{Cholesky decomposition:} Formulating the factorisation as the Cholesky decomposition of a banded infinite Toeplitz matrix. Herewith, we are avoiding the polynomial root solving entirely, providing numerical stability.
    \item \textit{Szeg\H{o}-Kolmogorov (Real Cepstrum) \cite{BarettNumerical1983}:} Evaluating the analytic projection in the frequency domain via the real cepstrum requires Fast Fourier Transformations. This completely bypasses matrix factorisations, allowing the construction of circuits with a high order of harmonics.
\end{itemize}

\section{Discussion}
We have presented a complete description of an optical Fourier encoding, capable of computing any finite Fourier series using exclusively $2\times 2$ linear optics and a fixed classical input state. We have proven that such a circuit always exists and requires solving a highly nonlinear optimisation program that nonetheless has an efficient solution method via classical spectral factorisation. This directly yields an efficient iterative procedure that constructs the required optical sequence out of simple fixed Mach-Zehnder Interferometers and tunable single-mode phase shifters, the latter of which encode the function input. For an $\mathcal{O}(N)$ term Fourier series, the resultant circuit has an asymptotic circuit depth that is bounded by $\mathcal{O}(N)$. Furthermore, the optical computation time is almost independent of the function's analytical complexity, governed purely by optical propagation delay through the circuit depth, and is practically dominated by the time required to set the tuneable phase shifters.

The algebraic reduction algorithm introduced here represents the optical analogue of the classical Schur algorithm on the unit disk, while providing an exact physical realisation of polynomial completion in Generalised Quantum Signal Processing \cite{motlaghGeneralizedQuantumSignal2024b}. By porting these algebraic insights into dual-rail classical coherent photonics, our framework demonstrates that techniques originally developed for quantum state manipulation can solve classical optical compilation bottlenecks.

While our results provide a complete solution for implementing one-dimensional nonlinear functions in linear optics, the underlying algebraic strategies used to do so are restricted to univariate polynomials. The natural extension to this work is whether, and how, one may do the same for multivariable functions $f(x,y)$. This would require a circuit synthesis over a Fourier series of the form
\begin{equation}
    f(x,y) = \sum_{s,t}w_{st}e^{i(sx + ty)},
\end{equation}
for which the Fej\'er-Riesz and spectral factorisation methods no longer guarantee a solution as in the single variable case. Alternatively, one might consider single-variable functions $f_2\circ f_1(x)$ and their composition on an initial argument $x$. While it would be exceedingly useful the output of an optical circuit is, unsurprisingly, optical in nature and using it to program further linear optical elements usually requires conversion into an electrical signal that is fed into the next circuit, thereby reintroducing an electro-optical bottleneck and reducing the effectiveness of chaining together such circuits. Realising optical circuits programmable by incoming optical light directly would bypass this conversion and unlock generalised, all-optical computation. Establishing which of these can be achieved using near-term classical optical platforms is not yet clear and serves as an important remaining open problem.

\acknowledgements
The authors acknowledge the use of generative AI assistance (in the form of ChatGPT-5.6 Sol and Antigravity powered by Gemini 3.8 Flash) for assistance in drafting and development of the simulation scripts; all algorithmic implementations, theoretical derivations, and numerical results were verified by and remain the sole responsibility of the authors. The numerical simulation scripts and circuit compilation routines developed in this work are openly available on Zenodo \cite{mauser_2026_23035668}. The computational results presented have been achieved in part using the Austrian Scientific Computing (ASC) infrastructure. This work was supported by the Austrian Research Promotion Agency (FFG) [PIQLearn ID: 54460501]. The presented work has been funded in part by the Vienna Science and Technology Fund (WWTF) [GrantID: 10.47379/ICT25062] (ALPAQA). This research was funded in part by the Austrian Science Fund (FWF)[10.55776/F71] (BeyondC). For open access purposes, the author has applied a CC BY public copyright license to any author-accepted manuscript version arising from this submission. Co-funded by the European Union (HORIZON Europe Research and Innovation Programme, EPIQUE, No 101135288). Views and opinions expressed are however those of the author(s) only and do not necessarily reflect those of the European Union or the European Commission-EU. Neither the European Union nor the granting authority can be held responsible for them.
\bibliography{bibliography}

\appendix
\section{Existence of linear optical circuit} \label{sec:existence}
\begin{thm}[Existence of a circuit]
    For any finite Fourier series of the form
    \begin{equation}
        f(x) = \sum_{n=p}^q a_n e^{ix n},
    \end{equation}
    there always exists an auxiliary polynomial $g(x)$ with the same support,
    \begin{equation}
        g(x) =  \sum_{n=p}^q b_n e^{ix n},
    \end{equation}
    such that
    \begin{equation}
        \label{eq:AppendixNormCondition}
        |f(z)|^2 + |g(z)|^2 = C, \qquad \forall |z|=1,
    \end{equation}
    for any constant $C > \max_{|z|=1} |f(z)|^2$, where $z=e^{ix}$.

\end{thm}
Define the target polynomial vector $F(z)$ as a  sum
\begin{equation}
	F(z)= \begin{pmatrix}
	    f(z)\\g(z)
	\end{pmatrix}=\sum_{n=p}^{q} z^n \ket{\phi_n},
	\quad
	\ket{\phi_n}\in\mathbb{C}^2,
	\quad
	|z|=1,
\end{equation}
with $\ket{\phi_n} := (a_n, b_n)^T$. 
We seek $g(z)$ such that $F(z)^\dagger F(z) =C$ for all $|z| = 1$.
Here $C$ is not universal but is chosen freely for each target, subject only to the lower bound 
\begin{equation}
    C> \max_{|z|=1}|f(z)|^2.
\end{equation}
Physically, $C$ is simply the total input power injected into the two modes, which we are free to select. The target function is recovered from the normalized output intensity, so this rescaling is required regardless.
Now, define a new strictly positive $S(z)$ such that
\begin{equation}
    S(z)=C-|f(z)|^2 > 0.
\end{equation}
The usual boundedness theorem from Fej\'er and Riesz gives us the following assertion for positive polynomials:
\begin{thm}[Fej\'er-Riesz]\cite{haagerup1992}
	If
	\[S(z)> 0 \qquad\forall z=e^{ix}, \]
	then there exists a polynomial
	\[ h(z)=\sum_{j=0}^{K} b_j z^j, \]
	where $K := q-p$ denotes the maximal lag, equal to the width of the Fourier support of $f$, such that
	\[ S(e^{ix})=|h(e^{ix})|^2,  \]
	for every $x$.
\end{thm}
The consequences of this are especially relevant here and give us a path to proving the existence of $g(z)$.
Given Eq.\eqref{eq:AppendixNormCondition}, set
\begin{equation}
	S(z)=C-|f(z)|^2. 
\end{equation}
Since $S(z) > 0$ by construction, Fejér--Riesz gives
\begin{equation}
	S(z)=|h(z)|^2
\end{equation}
on ($|z|=1$), with
\begin{equation}
	h(z)=b_0+b_1z+\cdots+b_Kz^K. 
\end{equation}
Now define a final polynomial that is $h(z)$ shifted by the monomial power $z^p$ as $g(z)=z^p h(z)$. Because $|z^p| = 1$, it remains that  $|g(z)|^2=|h(z)|^2$ and we have
\begin{equation}
	g(z) = z^p \sum_{j=0}^{K} b_j z^j = \sum_{n=p}^{q} b_{n-p} z^{n}
\end{equation}
so our $g(z)$ has the same series support as $f(z)$ and 
\begin{align*}
	|f(z)|^2 + |g(z)|^2  & = |f(z)|^2 + |z^p h(z)|^2,\\
                         & = |f(z)|^2 -|f(z)|^2 + C,\\
                         & = C.
\end{align*}
as required, settling the existence question in the affirmative. But what of the actual coefficients $b_n$? We know they must exist, but how are they computed given $f(z)$ in practice? The trick is to consider the set of coefficients globally rather than trying to assign them locally as part of the coming iterative procedure. As an intermediate step, we compute $S(z)$. It is easy to see when grouping by powers that
\begin{equation}
	|f(z)|^2 = \sum_{m,n} a_m^* a_n z^{n-m} = \sum_{r=-K}^K A_r z^r.
\end{equation}
for the autocorrelations $A_r = \sum_{n=p}^{q-r} a_n^* a_{n+r}$. Using the identities $A_0 = \sum_{n=p}^q |a_n|^2$ and $A_{-r} = A_r^*$ we see that 
\begin{equation}
	S(z) = C - |f(z)|^2 = \sum_{r=-K}^K c_r z^r
\end{equation}
with shifted coefficients $c_0 = C - \sum_{n=p}^{q} |a_n|^2$ and more generally
\begin{equation}
	c_{r \neq 0} = -A_r = - \sum_{n=p}^{q-r} a_n^* a_{n+r}.
\end{equation}
So we may compute $S(z)$ directly from the terms of $f(z)$. The next step is, given $S(z)=|h(z)|^2$, what is $h(z)$? The same trick applies with a bit more difficulty in the solution. If
\begin{equation}
	h(z)=\sum_{j=0}^{K} b_j z^j,
\end{equation}
then
\begin{equation}
	|h(z)|^2 = \sum_{j,\ell=0}^{K} b_j^* b_\ell z^{\ell - j} = \sum_{r=-K}^{K} \sum_{j=0}^{K-|r|} b_j^* b_{j+|r|} z^{r} = \sum_{r=-K}^K c_r z^r,
\end{equation}
thus, matching coefficients for $r \ge 0$ (negative lags follow from $c_{-r} = c_r^*$),
\begin{equation}
	c_{r}  = \sum_{j=0}^{K-r} b_j^* b_{j+r}, \quad r\in [0,K],
\end{equation}
and hence
\begin{equation}
	\sum_{j=0}^{K-r} b_j^* b_{j+r} = -\sum_{n=p}^{q-r} a_n^* a_{n+r} = -A_r, \quad r\neq 0,
\end{equation}
and that this set of equations must always have a solution in the $b_n$ given $a_n$. This gives a minimisation constraint in terms of the sum of collected $r$ modes and the unknown $b_n$; a satisfying solution yields 
\begin{equation}
	A_r + \sum_{j=0}^{K-r}  b_j^* b_{j+r}  =  0, \qquad r=1, \dots, K.
\end{equation}
The $r=0$ component is enforced by the normalisation constant $C$ through
\begin{equation}
	\mathcal{L}_0 = \sum_{j=0}^K b_j^* b_j + \sum_{j=p}^q a_j^* a_j - C = 0
\end{equation}
If we then optimise a vector $b \in \mathbb{C}^{K+1}$ over this constraint, we arrive at the optimisation objective that encapsulates the main difficulty with determining the auxiliary polynomial
\begin{equation}\label{eq:nonlinear_optimisation_problem}
 \min_b	\mathcal{L}(b) = \sum_{r=1}^K \left|A_r + \sum_{j=0}^{K-r}  b_j^* b_{j+r}\right|^2 + |\mathcal{L}_0|^2.
\end{equation}

Solving this yields the required coefficients ${b_n}$ for the auxiliary polynomial whose existence is now assured for any $f(z)$ and consequently for any finite Fourier series. This completes the existence proof.

\section{Spectral factorisation of auxiliary polynomial} \label{sec:root-finding}
The nonlinear optimisation problem presented in Eq.\eqref{eq:nonlinear_optimisation_problem} is a common outcome of the application of the Fej\'{e}r-Riesz theorem, and analytical methods for finding the roots of $h(z)$ and thus the coefficients $b_n$ of the auxiliary polynomial $g(z)$ are in abundance \cite{golubMatrixComputations2013, oppenheimDiscreteTimeSignalProcessing, BarettNumerical1983}. While the minimisation objective presented in Eq.\eqref{eq:nonlinear_optimisation_problem} is valid, it is not useful in practical terms as finding a global minimum over a sum of many polynomial equations consisting of more than a dozen terms is unlikely to succeed naively. Since the circuit construction cannot proceed without knowing the auxiliary polynomial, we elect to use one that is stable under polynomials with thousands of terms. In such cases, root-finding algorithms lack robustness, especially if any of the roots lie close to the unit circle, which leads to poor conditioning of eigenvalues that root-finding methods tend to rely on. We will give a brief outline of the method here based on repeated Cholesky decomposition of submatrices of an infinite Toeplitz matrix; for a complete description on why this approach yields the $b_n$ coefficients, see \cite{kolev2024}. From the previous section, we have the autocorrelation constraint on the collected $r$ harmonic modes of the Fourier series
\begin{equation}
	c_{r \neq 0} = - \sum_{n=p}^{q-r} a_n^* a_{n+r}.
\end{equation}
which we use to form a structured matrix whose entries are formed by the collected Fourier mode coefficients $c_r$. These encapsulate the orthogonality constraints that define a unique $g(z)$ and whose satisfaction is achieved by finding the global minimum $\mathcal{L}(b) = 0$ for the unknown coefficients $b_n$ in Eq.\eqref{eq:nonlinear_optimisation_problem}. Instead, we form the banded Toeplitz matrix with repeating structure
\begin{equation}
    \mathcal{T} = 
    \begin{pmatrix}
        c_0    & c_1^*    & c_2^*    & \cdots &  c_{K}^*   & 0          & \cdots \\
        c_1    & c_0      & c_1^*    & \cdots &  c_{K-1}^* & c_{K}^*   & \ddots \\
        c_2    & c_1      & c_0      & \cdots &  c_{K-2}^* & c_{K-1}^* & \ddots \\
        \vdots & \vdots   & \ddots   & \ddots &  \ddots     & \ddots     & \ddots \\
        c_{K} & c_{K-1} & \cdots   & \cdots &  c_{0}      & c_1^*      & \ddots \\
        0      & c_{K}   & c_{K-1} & \cdots &  c_{1}      & c_0        & \ddots \\
        \vdots &  \ddots  & \ddots   & \cdots &  \ddots     & \ddots     & \ddots \\
    \end{pmatrix}
\end{equation}
with $\mathcal{T}_{ij} = c_{i-j}$ for $|i-j|\leq K$. Note that this matrix is Hermitian as constructed and each row may have at most $2K+1$ nonzero entries, with the repeating shifted row structure extending to infinity. If we consider an $N \times N$ principal submatrix
\begin{equation}
    \mathcal{T}_m = [c_{i-j}]_{i,j=0}^{m-1}, \quad m \gg K,
\end{equation}
we may compute its Cholesky factorisation $\mathcal{T}_m = L_m L^\dagger_m$.

Positive definiteness of every principle section follows directly from the strict positivity of $S$: for any $v \in \mathbb{C}^m$ with associated polynomial $V(z) = \sum_{j=0}^{m-1} v_j z^j$,
\begin{equation}
    v^\dagger \mathcal{T}_m v
    = \frac{1}{2\pi}\int_0^{2\pi} S(e^{ix})\,|V(e^{ix})|^2\, \mathrm{d}x
    > 0,
\end{equation}
so the Cholesky decomposition is always well defined.

Additionally, since $\mathcal{T}$ is banded, the triangular matrices can be computed in banded form as well and do not require instantiation of the complete $m \times m$ matrix, speeding up the factorisation. The relevance of this here is Bauer's finite-selection result \cite{kolev2024}, which asserts that away from the matrix boundary the Cholesky factor satisfies
\begin{equation} 
b_k = \lim_{m\rightarrow\infty} (L_m)_{m-1,m-1-k}, \quad k=0, \ldots, K, 
\end{equation}
or, in other words, the final rows of $L_m$ converge to the desired polynomial coefficients. Intuitively, the factorisation $\mathcal{T} = L L^\dagger$ of the infinite matric into banded lower-trianglar Toeplitz factors is exactly the matrix form of the polynomial identity $S(z) = |h(z)|^2$. Finite sections inherit this structure away from the boundary, and because the minimum-phase factor has all roots strictly inside the unit circle, boundary effects decay exponentially quickly with row index \cite{kolev2024}. Indeed, the $r$th diagonal of the product $L_m L_m^\dagger$ approaches 
\begin{equation} 
\sum_{j=0}^{2K-r} b_j^*b_{j+r},
\end{equation} 
which is exactly the orthogonality condition required of the auxiliary polynomial. This is especially favourable numerically as factorisation over positive definite matrices tends to be much more stable than numerical root-finding-based methods and remains extremely fast, even for large principle sections, with cost set by the bandwidth $K$. As an example, the thousand-term auxiliary polynomial for Figure \ref{fig:main_wide}b was computed in approximately 1 second on a laptop equipped with an AMD Ryzen 7040 Series CPU. We thus have a robust method for finding the auxiliary polynomial $g(z)$ given $f(z)$, greatly simplifying the task of determining an optical circuit that generates the required unitary transformation on the vector $F(z)$.

\section{Iterative construction of inverse circuit} \label{sec:constrcution}
Having described a robust method of computing the coefficients $b_n$ of the required auxiliary polynomial 
\begin{equation}
	g(z)=\sum_{n=p}^{q} b_n z^n,
\end{equation}
such that
\begin{equation}\label{eq:constant_norm_completion}
	|f(z)|^2+|g(z)|^2=C,\qquad |z|=1,
\end{equation}
we now construct a unitary circuit which successively reduces the Fourier support. Since the coefficients \(b_n\) are now fixed globally, no further auxiliary parameters need to be chosen during the circuit construction. Let
\begin{equation}
	f(z)=\sum_{n=p}^{q}a_n z^n,	\qquad	g(z)=\sum_{n=p}^{q}b_n z^n,
\end{equation}
and define the coefficient vectors
\begin{equation}\label{eq:phi_definition}
	\ket{\phi_n} :=	a_n\ket{0}+b_n\ket{1} =
    \begin{pmatrix}
		a_n\\
		b_n
	\end{pmatrix}.
\end{equation}
The completed vector-valued polynomial is therefore
\begin{equation}\label{eq:completed_polynomial}
	F(z) :=
    \begin{pmatrix}
		f(z)\\
		g(z)
	\end{pmatrix}
	= \sum_{n=p}^{q}z^n\ket{\phi_n}.
\end{equation}
For each integer $r$, define the $r$-sums we used before
\begin{align}
	A_r	& := \sum_n a_n^*a_{n+r},	\label{eq:A_r_def}\\
	B_r	& := \sum_n b_n^*b_{n+r},	\label{eq:B_r_def}
\end{align}
where all coefficients outside the interval $[p,q]$ vanish. Expanding the norm of Eq.\eqref{eq:completed_polynomial} gives
\begin{equation}
	F(z)^\dagger F(z) =	\sum_{r=-(q-p)}^{q-p} \left( A_r+B_r \right) z^r.
\end{equation}
By construction, Eq.~\eqref{eq:constant_norm_completion} holds for every
\(|z|=1\). Since the collected Fourier modes $e^{irx}$ are orthogonal over one period, equality to a constant for every $x$ requires all the nonzero Fourier coefficients to vanish
\begin{equation}\label{eq:autocorrelation_constraints}
	A_r+B_r = \sum_n \braket{\phi_n|\phi_{n+r}} = C \delta_{r0},
\end{equation}
The maximal-lag condition immediately gives the endpoint orthogonality
required for the circuit construction. Indeed, for $r=q-p,$ there is only one pair of coefficients separated by this lag, namely $\phi_p$ and $\phi_q$. Hence
\begin{equation}
	A_{q-p}+B_{q-p}	= a_p^*a_q+b_p^*b_q	= \braket{\phi_p|\phi_q}.
\end{equation}
Since $q-p \neq 0 $, Eq.\eqref{eq:autocorrelation_constraints} yields our endpoint orthogonality condition
\begin{equation}\label{eq:endpoint_orthogonality}
	\braket{\phi_p|\phi_q}=0.
\end{equation}
Thus the Fej\'{e}r--Riesz completion supplies through these autocorrelation constraints the endpoint orthogonality needed at each stage of the reduction. Assume first that $\ket{\phi_p}$ and $\ket{\phi_q}$ are nonzero and define
\begin{equation}\label{eq:u_basis}
	\ket{u_0} = \frac{\ket{\phi_q}}{\norm{\phi_q}},\qquad
    \ket{u_1} = \frac{\ket{\phi_p}}{\norm{\phi_p}}.
\end{equation}
By Eq.\eqref{eq:endpoint_orthogonality}, these vectors form an orthonormal basis of $\mathbb{C}^2$. Define
\begin{equation}\label{eq:U_definition}
	U=\ket{0}\!\bra{u_0} + \ket{1}\!\bra{u_1},
\end{equation}
and a phase shifter that encodes the argument $T(z) = z^{-1}\outerproduct{0} +	\outerproduct{1}$. Note that this is the inverse of the physical phase shifter of Eq.~\eqref{eq:phase_gate}; the physical circuit is recovered upon inversion of the equence in Eq.~\eqref{eq:final_circuit}. Then
\begin{equation}\label{eq:boundary_conditions}
	\braket{u_0|\phi_p}=0, \qquad \braket{u_1|\phi_q}=0.
\end{equation}
Composing the iterative step as the joint operation $G(z):=T(z)U$, the state evolves as $F'(z):=G(z)F(z)$ and we obtain
\begin{align}\label{eq:one_step_expansion}
	F'(z) &= \sum_{n=p}^{q}	z^n	\left(z^{-1}\braket{u_0|\phi_n}\ket{0} + \braket{u_1|\phi_n}\ket{1} \right)	\nonumber \\
	       &= z^{p-1} \braket{u_0|\phi_p}\ket{0} + \sum_{n=p}^{q-1}z^n\ket{\phi_n'} + z^q \braket{u_1|\phi_q}\ket{1},
\end{align}
where
\begin{equation}\label{eq:coefficient_recurrence}
	\ket{\phi_n'} =	\braket{u_0|\phi_{n+1}}\ket{0} + \braket{u_1|\phi_n}\ket{1}.
\end{equation}
The two boundary terms in Eq.\eqref{eq:one_step_expansion} vanish with Eq.\eqref{eq:boundary_conditions}, and therefore the support of the new series is decremented $[p,q] \rightarrow [p, q-1]$. Moreover, the left endpoint remains nonzero whenever $\ket{\phi_p}\neq0$, since
\begin{equation}
	\braket{u_1|\phi_p} = \norm{\phi_p}.
\end{equation}
Thus the reduction preserves the left endpoint $p$, while decreasing the right endpoint by at least one. For $|z|=1$, both $T(z)$ and $U$ are unitary. Consequently,
\begin{equation}\label{eq:norm_preservation}
	F'(z)^\dagger F'(z)	= F(z)^\dagger F(z)	= C.
\end{equation}
Hence the constant-norm condition, and therefore the complete family of autocorrelation constraints
\begin{equation}
	A_r'+B_r'=C\delta_{r0},
\end{equation}
is preserved after the transformation. In particular, the maximal-lag relation for the reduced polynomial once more implies the orthogonality of the newly active endpoint coefficient vectors, allowing the same construction to be applied iteratively. Let
\begin{equation}
	F_0(z)= \sum_{n=p}^{q}z^n\ket{\phi_n^{(0)}},
\end{equation}
denote the initial completed polynomial which satisfies
\begin{equation}
	F_0(z)^\dagger F_0(z)=C.
\end{equation}
At the $k$-th reduction step, define
\begin{equation}
	F_k(z) = G_k(z)F_{k-1}(z),	\qquad G_k(z)=T(z)U_k.
\end{equation}
Since every $G_k(z)$ is unitary on the unit circle
\begin{equation}
	F_k(z)^\dagger F_k(z)=C,
\end{equation}
at every stage. Therefore, the corresponding mode groupings satisfy
\begin{equation}
	A_r^{(k)}+B_r^{(k)} = C\delta_{r0},
\end{equation}
at every iterative step; thus, the maximal nonzero lag again provides the endpoint orthogonality
required to construct $U_{k+1}$ every time. Each step decreases the series width by one while preserving
the left endpoint. Hence, after $L \leq q-p$ steps, only the coefficients $F_L(z) = z^p\ket{c}$ remains
for some constant vector $\ket{c}\in\mathbb{C}^2$ up to the global phase $z^p$. Since
\begin{equation}
	F_L(z) = G_L(z)G_{L-1}(z)\cdots G_1(z)F_0(z),
\end{equation}
reversing the sequence gives
\begin{equation}\label{eq:final_circuit}
	F_0(z) = G_1(z)^\dagger G_2(z)^\dagger \cdots G_L(z)^\dagger z^p\ket{c},
\end{equation}
and the amplitude in the first mode is proportional to the target polynomial $f(z)$, completing the circuit construction.

\section{Numerical Comparison of Spectral factorisation Algorithms} \label{sec:numerical_comp}

As described above, the deterministic compilation of the optical Fourier encoding circuit of a target function $f(z)$ requires generating an auxiliary polynomial $g(z) = \sum_{n=-N}^N b_n z^n$ such that $|f(z)|^2 + |g(z)|^2 = C$ holds. While our primary circuit synthesis adopts banded Cholesky decomposition on the Toeplitz matrix, we systematically evaluated its performance against the algebraic root-pairing companion eigensolver \cite{oppenheimDiscreteTimeSignalProcessing,golubMatrixComputations2013} and the Szeg\H{o}-Kolmogorov real-cepstrum algorithm \cite{BarettNumerical1983} across Fourier cutoffs spanning a domain from $K=5$ to $K=1000$ (corresponding to $2001$ active Fourier harmonics).

\begin{figure*}
    \centering
    \includegraphics[width=0.99\linewidth]{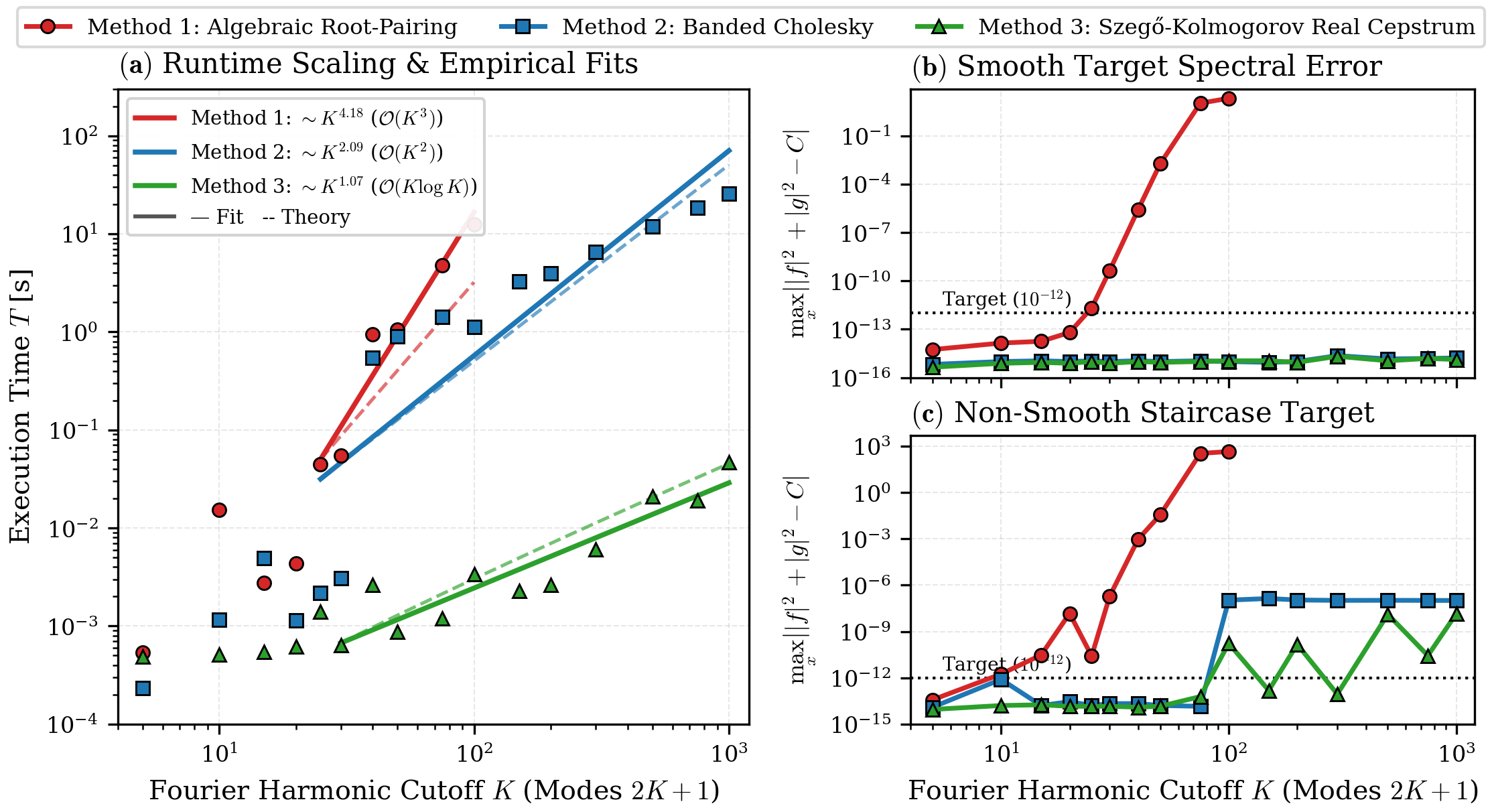}
    \caption{\textbf{Comprehensive benchmark of spectral factorisation algorithms.} \textit{(a)} Execution runtime scaling on a log-log scale. Empirical benchmark measurements (markers) are fitted to asymptotic power laws $T(K) \propto K^\alpha$ (solid lines), with respect to theoretical runtime expectation (dashed lines). Hereby, the algebraic root-pairing scales as $\sim K^{4.18}$ before diverging, while the banded Cholesky and Szeg\H{o}-Kolmogorov scale with $\sim K^{2.09}$ and $\sim K^{1.07}$ respectively, closely matching their theoretical scaling. \textit{(b)} Spectral unitary residual $\| C - |f|^2 - |g|^2\|_\infty$ on a smooth Gaussian-mixture target function. Both Cholesky and Szeg\H{o}-Kolmogorov sustain uniform machine-precision accuracy across all examined orders, whereas root-pairing undergoes catastrophic ill-conditioning beyond $K \approx25$, diverging exponentially thereafter. \textit{(c)} Spectral unitary residual on a discontinuous staircase target featuring severe Gibbs ringing. Both Cholesky and Szeg\H{o}-Kolmogorov exhibit numerical robustness in the examined domain, with the spectral errors being strictly bounded by $10^{-6}$. In contrast, root-pairing breaks down due to ill-conditioning, with residuals exceeding $10^2$.}
    \label{fig:factorisation_methods}
\end{figure*}

The three algorithms exhibit markedly different scaling and stability regimes (Fig.~\ref{fig:factorisation_methods}). Algebraic root-pairing becomes numerically unusable beyond $K \ge 25$, where dense clustering of high-degree roots near $|z|=1$ corrupts the minimum-phase partition under finite-precision arithmetic, causing exponential error divergence. In contrast, banded Cholesky and Szeg\H{o}-Kolmogorov maintain flat machine precision ($\sim 10^{-15}$) for smooth functions (Fig~\ref{fig:factorisation_methods}b), and remain stably bounded ($<10^{-6}$) even under severe Gibbs oscillations (Fig~\ref{fig:factorisation_methods}c).

\end{document}